# Intrinsic nature of chiral charge order in the kagome superconductor RbV$_3$Sb$_5$

**Authors:** Nana Shumiya[1]*, Md Shafayat Hossain[1]*, Jia-Xin Yin[1]*†, Yu-Xiao Jiang[1]*, Brenden R. Ortiz[2], Hongxiong Liu[3,4], Youguo Shi[3,4], Qiangwei Yin[5], Hechang Lei[5], Songtian S. Zhang[1], Guoqing Chang[6], Qi Zhang[1], Tyler A. Cochran[1], Daniel Multer[1], Maksim Litskevich[1], Zi-Jia Cheng[1], Xian P. Yang[1], Zurab Guguchia[7], Stephen D. Wilson[2], M. Zahid Hasan[1,8,9,10]†

**Affiliations:**

[1]Laboratory for Topological Quantum Matter and Advanced Spectroscopy (B7), Department of Physics, Princeton University, Princeton, New Jersey, USA.

[2]Materials Department and California Nanosystems Institute, University of California Santa Barbara, Santa Barbara, California 93106, USA.

[3]Beijing National Laboratory for Condensed Matter Physics and Institute of Physics, Chinese Academy of Sciences, Beijing 100190, China.

[4]University of Chinese Academy of Sciences, Beijing 100049, China.

[5]Department of Physics and Beijing Key Laboratory of Opto-electronic Functional Materials & Micro-nano Devices, Renmin University of China, Beijing 100872, China.

[6]Division of Physics and Applied Physics, School of Physical and Mathematical Sciences, Nanyang Technological University, Singapore 637371, Singapore.

[7]Laboratory for Muon Spin Spectroscopy, Paul Scherrer Institute, CH-5232 Villigen PSI, Switzerland.

[8]Princeton Institute for the Science and Technology of Materials, Princeton University, Princeton, New Jersey 08544, USA.

[9]Materials Science Division, Lawrence Berkeley National Laboratory, Berkeley, California 94720, USA.

[10]Quantum Science Center, Oak Ridge, Tennessee 37831, USA.

†Corresponding authors, E-mail: jiaxiny@princeton.edu; mzhasan@princeton.edu
*These authors contributed equally to this work.

**Superconductors with kagome lattices have been identified for over 40 years, with a superconducting transition temperature T$_C$ up to 7K. Recently, certain kagome superconductors have been found to exhibit an exotic charge order, which intertwines with superconductivity and persists to a temperature being one order of magnitude higher than T$_C$. In this work, we use scanning tunneling microscopy (STM) to study the charge order in kagome superconductor RbV$_3$Sb$_5$. We observe both a 2×2 chiral charge order and nematic surface superlattices (predominantly 1×4). We find that the 2×2 charge order exhibits intrinsic chirality with magnetic field tunability. Defects can scatter electrons to introduce standing waves, which couple with the charge order to cause extrinsic effects. While the chiral charge order resembles that discovered in KV$_3$Sb$_5$, it further interacts with the nematic surface superlattices that are absent in KV$_3$Sb$_5$ but exist in CsV$_3$Sb$_5$.**

Kagome lattices [1], made of corner sharing triangles, are tantalizing quantum platforms for studying the interplay between geometry, topology, and correlation. For instance, insulating kagome magnets have been investigated for decades in the hopes of realizing quantum spin liquids [2]. Recently, focused STM research on correlated kagome magnets has revealed many topological and many-body phenomena [3], including Chern gapped phases [4][5], tunable electronic nematicity [4], orbital magnetism [6,7,8], and many-body interplay [9][10]. These observations are all closely related to the emergent physics arising from the fundamental kagome band structure, which includes Dirac cones, flat bands, and van Hove singularities. Notably, the many-body fermion-boson interplay [10] observed in certain kagome paramagnets leads us to conjecture from the spectroscopic point of view that there can be superconductivity instability that competes with magnetism. Then, we realize that kagome superconductors with competing magnetism have been identified for at least over 40 years [11], such as $LaRu_3Si_2$ with $T_C$ of 7K and a fundamental kagome band structure [12]. Recently, another layered kagome superconductor $AV_3Sb_5$ (A = K, Rb, Cs) was discovered [13,14,15], providing new research opportunities, particularly for STM studies [16,17,18,19,20,21]. While $RbV_3Sb_5$ has been studied by several experimental techniques [15,22,23], it has not yet been studied with STM. In our earlier studies [16], we have reported the chiral 2×2 charge order in $KV_3Sb_5$, which displays robust chirality with magnetic field tunability on the defect-free region. Now, we find that $RbV_3Sb_5$ also features 2×2 charge order with additional nematic superlattices. It is crucial to reconfirm the chiral charge order in this new material and test its robustness against the surface superlattices.

$RbV_3Sb_5$ has a layered structure with the stacking of $Rb_1$ hexagonal lattice, $Sb_2$ honeycomb lattice, $V_3Sb_1$ kagome lattice, and $Sb_2$ honeycomb lattice shown in Fig. 1(a)-(c). Owing to the bonding length and geometry, the V and Sb layers have a stronger chemical bonding, and the material tends to cleave between Rb and Sb layers. The Sb surface is most interesting, as it is strongly bonded to the V kagome lattice. Previous STM studies have unambiguously resolved Sb honeycomb surfaces in $KV_3Sb_5$ and $CsV_3Sb_5$ [16,17,18,19,20,21], and studied the charge order and surface superlattices. We study $RbV_3Sb_5$ with STM at 4.2K. Through cryogenic cleaving, we have also obtained large clean Sb surfaces in $RbV_3Sb_5$, as shown in Fig. 1(d). The Fourier transform of the topography reveals a 2×2 charge order as marked by the shaded red region in Fig. 1(e). Such 2×2 charge order has been consistently observed in $KV_3Sb_5$ and $CsV_3Sb_5$ by both STM [15,16,17,18,19,21] and bulk [13,16] X-ray measurements. In addition, there are also nematic superlattice modulations (predominantly 1×4) along the $Q_1$ direction, and other weaker superlattice signals along this direction. A similar superlattice signal is also observed in the Sb surface in $CsV_3Sb_5$ [17,18,19,21]. However, such a signal has not been detected in K/Cs/Rb surfaces (or in the bulk X-ray data), while a bulk modulation will project and appear on all surfaces. Therefore, while the nematicity may be a bulk phenomenon, the specific 1×4 modulation is more likely to be a surface phenomenon. $KV_3Sb_5$ does not exhibit a 1×4 superlattice for Sb surface [16], and because $CsV_3Sb_5$ has a factor of three higher $T_C$ than that of $KV_3Sb_5$, previous STM observations in $CsV_3Sb_5$ conjectured a close relationship between the 1×4 superlattices and higher $T_C$ [19]. Our new observation in $RbV_3Sb_5$, which has a $T_C$ similar to $KV_3Sb_5$, makes such a scenario unlikely.

Now we focus on the intrinsic anisotropy of the 2×2 charge order on a large defect-free region in Fig. 2. We perform spectroscopic dI/dV maps at the same region with a magnetic field perpendicular to the surface. The maps taken at 30meV with B = 0T, -3T, +3T are displayed in Figs. 2(a)-(c), respectively. We find that the +3T map is different from the others. To better visualize the difference, we perform Fourier transform analysis of these maps. Particularly, we extract the six 2×2 vector peaks as shown in Figs. 2(d)-(f), which reveals pronounced intensity anisotropy along with different directions for all cases. This corresponds to the fact that the amplitudes of the 2×2 modulation in real-space along three directions are different with each other. The observed anisotropy can be due to a chiral charge order as initially discussed in certain transition-metal dichalcogenides and high-temperature superconductors [24,25]. The chirality can be defined as the counting direction (clockwise or anticlockwise) from the lowest to highest vector peaks. We find the chirality at the same atomic area can be switched by the magnetic field applied along the c-axis for opposite directions. A real space elaboration of the chirality switch is further shown in Fig. 3, demonstrating that the strength of 2×2 modulation is switched by the magnetic field. Figure 4 further shows the energy-resolved vector peak intensity for different magnetic fields. The vector peaks have weak intensity for negative energies, hindering the identification of chirality. For higher positive energies where the intensities are strong, we observe strong anisotropy. The intensity of $Q_1$ is always the strongest, and we note that this direction is the same as that of the nematic superlattices. Moreover, the intensities between $Q_2$ and $Q_3$ are different, from which we can determine chirality. The reversal of their intensities between -3T and +3T then demonstrates a chirality switch.

As a comparison, we also perform experiments around the defect-rich region in Fig. 5. Defects can backscatter electrons to induce standing waves. Figure. 4(a) shows rich standing waves in the dI/dV map of this region. The Fourier transform of this map shows clear ring-like signals just within the 2×2 charge order vector peaks. A detailed plot of the energy-resolved vector peak intensity at B = 0T, -3T, +3T is displayed in Fig. 4(b). Different from the defect-free case in Figs. 2-4, $Q_2$ and $Q_3$ basically have similar intensities over all measured energies, suggesting a diminishing of chirality. Moreover, there is no strong magnetic field response. All these observations are again consistent with our reports for $KV_3Sb_5$ [16]. We believe, because the standing wave signals in the q-space are close to the 2×2 charge order peaks, there exists a defect-pinning effect [26], which is an extrinsic property of the charge order. The interplay between the charge order and defects can be studied by Bogoliubov-de Gennes method in future.

Now we discuss the implications of our experiments. The new observations not only reconfirm the ubiquitous chiral charge order in $AV_3Sb_5$, but also suggest that the chirality and field switching are both robust against nematic superlattices. The 2×2 charge order has been proposed by pioneering theories of kagome lattices [27,28,29] at van Hove singularity filling. Recently, several theoretical works focused on $AV_3Sb_5$ [16,30,31,32,33,34,35,36] have confirmed 2×2 charge order with unconventional features, including time-reversal symmetry breaking, chirality, nematicity, and topology. The unconventional features arise from the interferences of three kagome sublattices with extended Coulomb interactions, and they can further interact with the topologically nontrivial band structure in these materials. While the nematicity of the charge order observed here can be consistent with the surface manifestation of the 2×2×2 charge order [37], the chirality ubiquitously observed in $KV_3Sb_5$ [16], $RbV_3Sb_5$ (this work) and $CsV_3Sb_5$ [38] cannot be explained by the conventional 2×2×2 charge order. As the chirality can be switched by a magnetic field that explicitly breaks time-reversal symmetry, it implies a complex set of order parameters of the charge order, which contain relative phase differences. The phase difference of three sets of the 2×2 order parameter, if not 0 or π, breaks time-reversal symmetry. Recently, more direct evidence of the time-reversal symmetry breaking comes from muon spin spectroscopy by observation of a concurrent emergence of an internal magnetic field with the charge order phase transition [39]. Theoretically, a broken time-reversal symmetry charge order is also suggested to be energetically favorable in the kagome lattice at van Hove filling and with extended Coulomb interactions [16,30,32,33,36], which features orbital currents running in the kagome lattice. Originally, charge order with broken time-reversal symmetry was proposed as the Haldane model for achieving quantum anomalous Hall effect [40] and orbital currents [41,42] for modeling pseudogap phase of cuprates. Its tantalizing visualization in kagome superconductors comes as an experimental surprise. Since there has not yet been anomalous Hall measurements for RbV$_3$Sb$_5$, whether our observed intrinsic and extrinsic behavior of the chiral charge order can be related with the intrinsic and extrinsic anomalous Hall effects [43,44] deserves future attention. It is also crucial to probe the magnetic field switching effect more systematically in future by varying the magnetic field strength, which can help to determine the critical switching field and to further compare with anomalous Hall measurements.

**Figures**

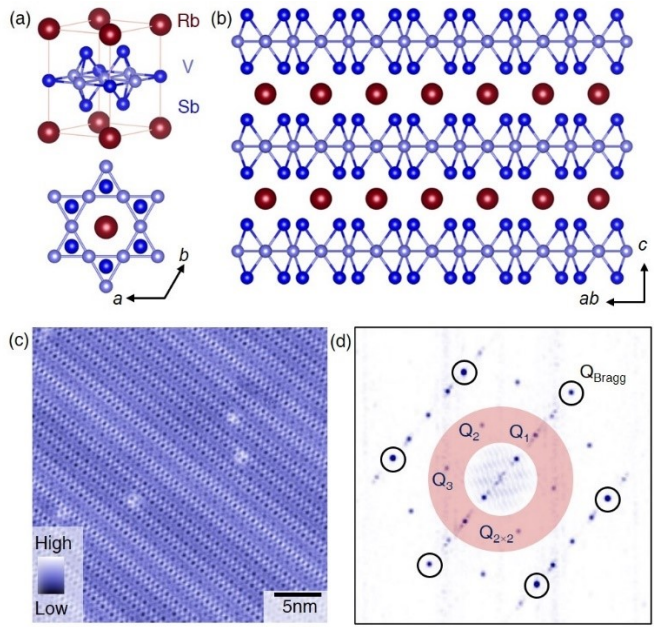

Fig. 1. (a)-(c) Crystal structure of kagome superconductor RbV$_3$Sb$_5$ from 3D view, top view, and side view, respectively. (d) Topographic image of a clean Sb surface (V = -100mV, I = 0.5nA). (e) Fourier transform of the topography showing Bragg peaks and charge ordering vector peaks. The 2×2 charge order vector peaks are highlighted by the shaded red ring, which includes pairs of Q$_1$, Q$_2$, and Q$_3$.

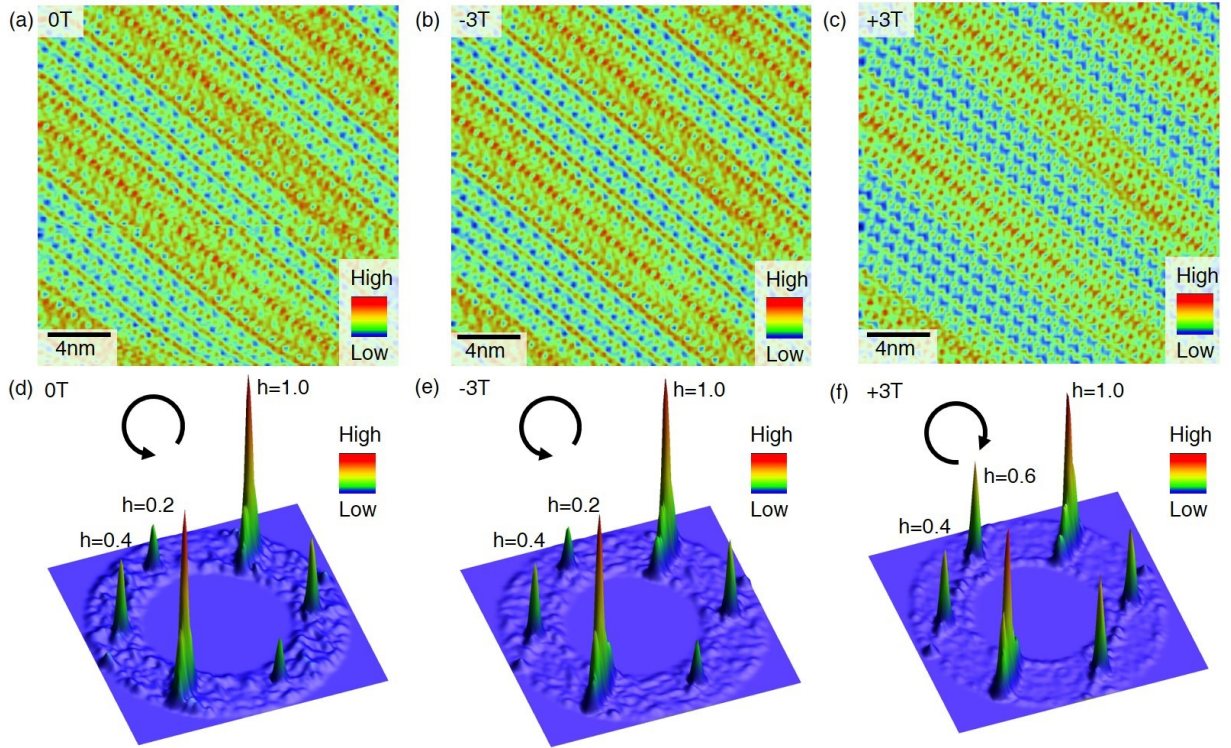

Fig. 2. (a)-(c) dI/dV maps taken at the same clean Sb surface with B = 0T, -3T, +3T, respectively. The magnetic field is applied along the c-axis. The maps are all taken at E = 30meV with V = -100mV and I = 0.5nA. (d)-(f) Spectroscopic 2×2 vector peaks taken at B = 0T, -3T, +3T, respectively. The images are Fourier transforms of spectroscopic maps. A circular region of the full Fourier-transformed image is shown for clarity, highlighting the six 2×2 vector peaks. The height of the three pairs of vector peaks is marked with arbitrary units for each data. The chirality can be defined as the counting direction (clockwise or anticlockwise) from the lowest to highest pair vector peaks as marked by the rotating arrows.

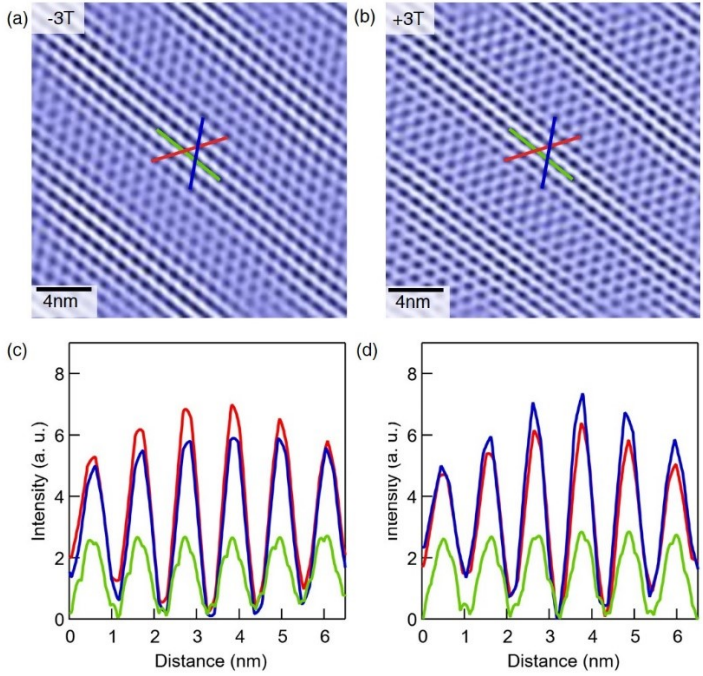

Fig. 3. (a)(b) Real space image of the 2×2 chiral charge order taken at B = -3T and B = +3T, respectively. The data is produced by the inverse Fourier transform of the 2×2 vector peaks in Fig. 2(e) and (f), respectively. (c)(d) Line-cut profile for along three directions marked in (a) and (b), respectively. The modulations along the three directions are different in both cases, defining a chirality. Magnetic field switch induces a switch of the strengths of two stronger modulations.

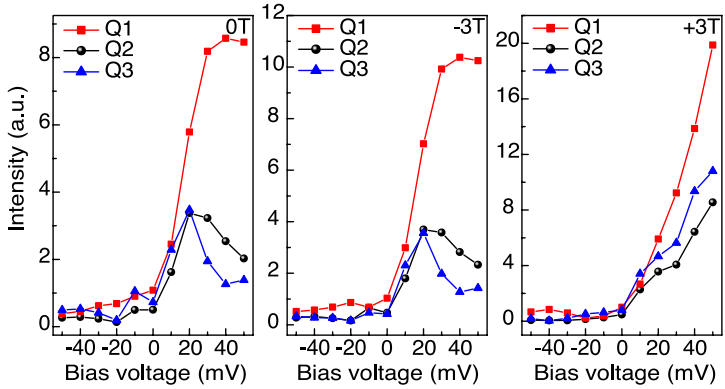

Fig. 4. Magnetic field tunability of the chiral charge order at a defect-free region. Comparison of intensities of three 2×2 vector peaks as a function of energy for the same defect-free region for B = 0T, -3T, +3T, respectively. The vector peaks have weak intensities at negative energies, and the magnetic field induced chirality switching effects are primarily observed at higher positive energies.

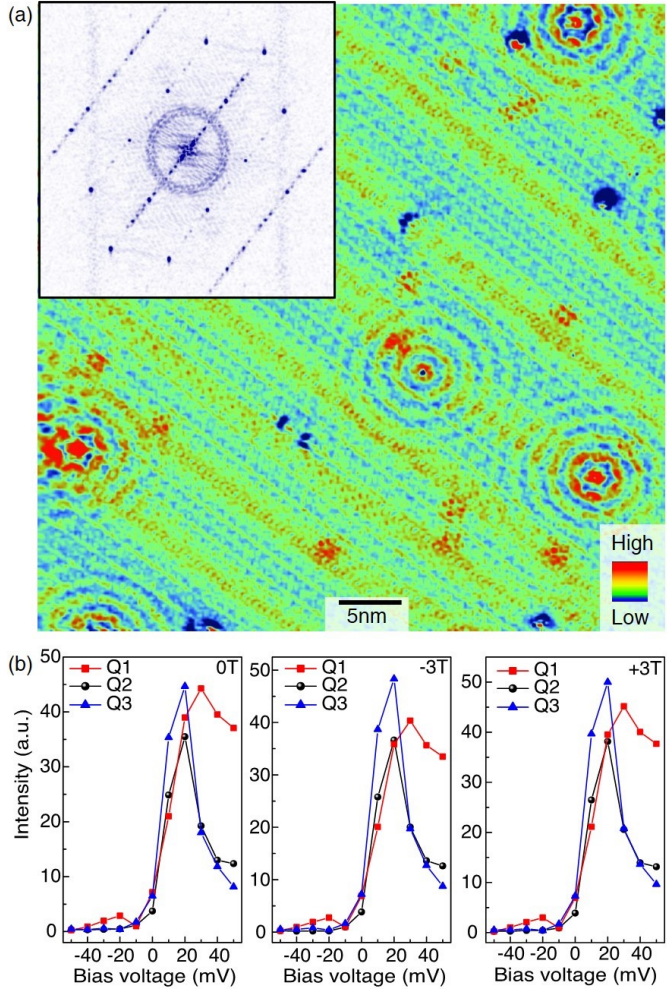

Fig. 5. Absence of chirality and magnetic tunability at standing-waves-rich region. (a) dI/dV map data taken at a defect-rich Sb surface. The maps are all taken at E = 0meV with V = -100mV and I = 0.5nA. This region hosts numerous defect-induced standing waves. The inset shows the Fourier transform of the map data, which exhibits additional ring-like signals within the 2×2 vector peaks. (b) Comparison of intensities of three 2×2 vector peaks for this defect-rich region as a function of energy for B = 0T, -3T, +3T, respectively. In this region, there is no apparent chirality of the charge order, and we do not observe a strong magnetic field response of the vector peaks.

**Acknowledgments**. Experimental and theoretical work at Princeton University was supported by the Gordon and Betty Moore Foundation (GBMF4547 and GBMF9461; M.Z.H.). The material characterization is supported by the United States Department of Energy (US DOE) under the Basic Energy Sciences programme (grant number DOE/BES DE-FG-02-05ER46200). S.D.W. and B.R.O. acknowledge support from the University of California Santa Barbara Quantum Foundry, funded by the National Science Foundation (NSF DMR-1906325). Research reported here also made use of shared facilities of the UCSB MRSEC (NSF DMR-1720256). B.R.O. also acknowledges support from the California NanoSystems Institute through the Elings fellowship program. T.A.C. was supported by the National Science Foundation Graduate Research Fellowship Program under Grant No. DGE-1656466. H.C.L. was supported by National Natural Science Foundation of China (Grant No. 11822412 and 11774423), Ministry of Science and Technology of China (Grant No. 2018YFE0202600 and 2016YFA0300504) and Beijing Natural Science Foundation (Grant No. Z200005). Y.S. was supported by the National Natural Science Foundation of China (U2032204), and the K. C. Wong Education Foundation (GJTD-2018-01). G.C. would like to acknowledge the support of the National Research Foundation, Singapore under its NRF Fellowship Award (NRF-NRFF13-2021-0010) and the Nanyang Assistant Professorship grant from Nanyang Technological University.